\documentclass[aps,prl,reprint,superscriptaddress]{revtex4-2}
\usepackage{graphicx}
\usepackage{amssymb}
\usepackage{amsmath}
\usepackage{mathrsfs}
\usepackage{color}
\usepackage{mathtools}
\usepackage[colorlinks,linkcolor=blue,anchorcolor=blue,citecolor=blue,urlcolor=blue]{hyperref}
\usepackage{physics}
\usepackage{bm}

\begin{document}

\title{Quantum Interference and Coherent Population Trapping in a Double Quantum Dot}

\author{Yuan Zhou}
\affiliation{CAS Key Laboratory of Quantum Information, University of Science and Technology of China, Hefei, Anhui 230026, China}
\affiliation{CAS Center for Excellence and Synergetic Innovation Center in Quantum Information and Quantum Physics, University of Science and Technology of China, Hefei, Anhui 230026, China}

\author{Ke Wang}
\affiliation{CAS Key Laboratory of Quantum Information, University of Science and Technology of China, Hefei, Anhui 230026, China}
\affiliation{CAS Center for Excellence and Synergetic Innovation Center in Quantum Information and Quantum Physics, University of Science and Technology of China, Hefei, Anhui 230026, China}

\author{He Liu}
\affiliation{CAS Key Laboratory of Quantum Information, University of Science and Technology of China, Hefei, Anhui 230026, China}
\affiliation{CAS Center for Excellence and Synergetic Innovation Center in Quantum Information and Quantum Physics, University of Science and Technology of China, Hefei, Anhui 230026, China}

\author{Gang Cao}
\affiliation{CAS Key Laboratory of Quantum Information, University of Science and Technology of China, Hefei, Anhui 230026, China}
\affiliation{CAS Center for Excellence and Synergetic Innovation Center in Quantum Information and Quantum Physics, University of Science and Technology of China, Hefei, Anhui 230026, China}

\author{Guang Can Guo}
\affiliation{CAS Key Laboratory of Quantum Information, University of Science and Technology of China, Hefei, Anhui 230026, China}
\affiliation{CAS Center for Excellence and Synergetic Innovation Center in Quantum Information and Quantum Physics, University of Science and Technology of China, Hefei, Anhui 230026, China}

\author{Xuedong Hu}
\affiliation{Department of Physics, University at Buffalo, SUNY, Buffalo, New York 14260, USA}

\author{Hai-Ou Li}
\email[]{haiouli@ustc.edu.cn}
\affiliation{CAS Key Laboratory of Quantum Information, University of Science and Technology of China, Hefei, Anhui 230026, China}
\affiliation{CAS Center for Excellence and Synergetic Innovation Center in Quantum Information and Quantum Physics, University of Science and Technology of China, Hefei, Anhui 230026, China}

\author{Guo-Ping Guo}
\email[]{gpguo@ustc.edu.cn}
\affiliation{CAS Key Laboratory of Quantum Information, University of Science and Technology of China, Hefei, Anhui 230026, China}
\affiliation{CAS Center for Excellence and Synergetic Innovation Center in Quantum Information and Quantum Physics, University of Science and Technology of China, Hefei, Anhui 230026, China}
\affiliation{Origin Quantum Computing Company Limited, Hefei, Anhui 230026, China}

\date{\today}

\begin{abstract}

Quantum interference is a natural consequence of wave-particle duality in quantum mechanics, and is widely observed at the atomic scale.  One interesting manifestation of quantum interference is coherent population trapping (CPT), first proposed in three-level driven atomic systems and observed in quantum optical experiments.  Here, we demonstrate CPT in a gate-defined semiconductor double quantum dot (DQD), with some unique twists as compared to the atomic systems.  Specifically, we observe CPT in both driven and non-driven situations.  We further show that CPT in a driven DQD could be used to generate adiabatic state transfer.  Moreover, our experiment reveals a non-trivial modulation to the CPT caused by the longitudinal driving field, yielding an odd-even effect and a tunable CPT.

\end{abstract}

\maketitle
\section{Introduction}

Quantum interference is one of the most vivid illustrations of the wave nature of quantum mechanical systems.  It shows up whenever multiple paths of different phases or multiple energy levels are present, from double slit experiments and weak localization to Landau-Zener interference.  Quantum interference also underlies many important applications, whether it is atomic clocks \cite{Degen2017} or Shor's factorization algorithm \cite{Shor1994, Ekert1996, Shor1997}, and would conceivably be a key feature to many future quantum coherent devices. 

One example of the consequences of quantum interference is coherent population trapping (CPT),  as a result of destructive interference between different transition paths, and first observed in a three-level atom in an optical experiment \cite{Gray1978}. 
In such a three-level system, two states are coupled to a third, intermediate, state.  When the driving fields are properly detuned for the two allowed transitions, a superposition of the first two states emerges and is decoupled from the intermediate states.
Such a superposition is called a ``dark state'' since a system trapped in this state would not respond to the probe field, leading to interesting phenomena such as electromagnetically induced transparency \cite{Harris1990}.
By adiabatically tuning the controls of the dark state (such as the amplitude and phase of the driving fields), one can perform rapid state initialization and Stimulated Raman Adiabatic Passage (STIRAP) with suppressed excitation \cite{Xu2007, Rogers2014, Gaubatz1990, Vitanov2017}.
Compared to state manipulation based on resonant driving, STIRAP has significant advantages in robustness, while maintaining high degree of efficiency and selectivity, and is thus of great importance in quantum information processing \cite{Gaubatz1990, Vitanov2017}.

CPT has been extensively studied in various physical systems since its first observation, and widely employed in both metrology and quantum engineering, from atomic cooling \cite{Aspect1988} and atomic clock \cite{Vanier2005}, to high-sensitivity magnetometry \cite{Scully1992, Nagel1998} and quantum state manipulation \cite{Gaubatz1990, Rogers2014, Vitanov2017}.  In condensed matter systems, CPT has been demonstrated in doped crystals \cite{Goto2007, Klein2008}, ultracold atoms \cite{Takekoshi2014, Molony2014}, color centers in diamond \cite{Golter2014}, superconducting circuits \cite{Kelly2010, Xu2016, Kumar2016}, microelectromechanical systems \cite{Knappe2008} and self-assembled quantum dots \cite{Xu2008, Brunner2009}.
In contrast, in electrically-controlled gate-defined quantum dots (QDs), which are viewed as one of the promising platforms for quantum computing \cite{Hendrickx2021, Adam2022, Noiri2022, Xue2022, Zwerver2022, Camenzind2022}, CPT has received less attention.  This is due to differences in energy structure and the way electromagnetic field couples to the system, even though QDs, particularly double quantum dots (DQD), provide intriguing possibilities for CPT.
For example, in an atom, the optical field participates in CPT in two aspects, to couple the energy levels, and to serve as a probe.
On the other hand, DQD energy levels are tunable, and its state can be readout via electronic methods, so that CPT without driving fields and CPT measurement without a probe field are both possible.

In this letter, we demonstrate CPT in a gate-defined DQD without drive or probe field, and also with one longitudinal drive.
We focus on a singlet-triplet (ST) system \cite{Petta2005, Pioro-Ladriere2008, Foletti2009,Huang2019, Hendrickx2020} in a DQD device for holes, and measure the leakage current through the DQD in the Pauli Spin Blockade (PSB) regime \cite{Kouwenhoven1997, Ono2002, Petta2005}.  In the absence of driving, we observe a sharp dip in the leakage current at zero bias, which can be attributed to the formation of dark states and the occurrence of CPT.
When the ST system is driven longitudinally through detuning $\varepsilon$, we again observe CPT under proper conditions, and the physical picture can be explained using an effective Hamiltonian \cite{Bukov2015, Oka2019, Weitenberg2021, Zhou2022} for this longitudinally driven system (with distinctive features compared to the transversely driven atomic systems, such as the modulation of the effective couplings and the observation of the odd-even effect \cite{Stehlik2014, Danon2014, Zhou2022}).  Importantly, with controls provided by the driving, we show that STIRAP is feasible.

\begin{figure}[tbp]
    \centering
    \includegraphics[width=8.6cm]{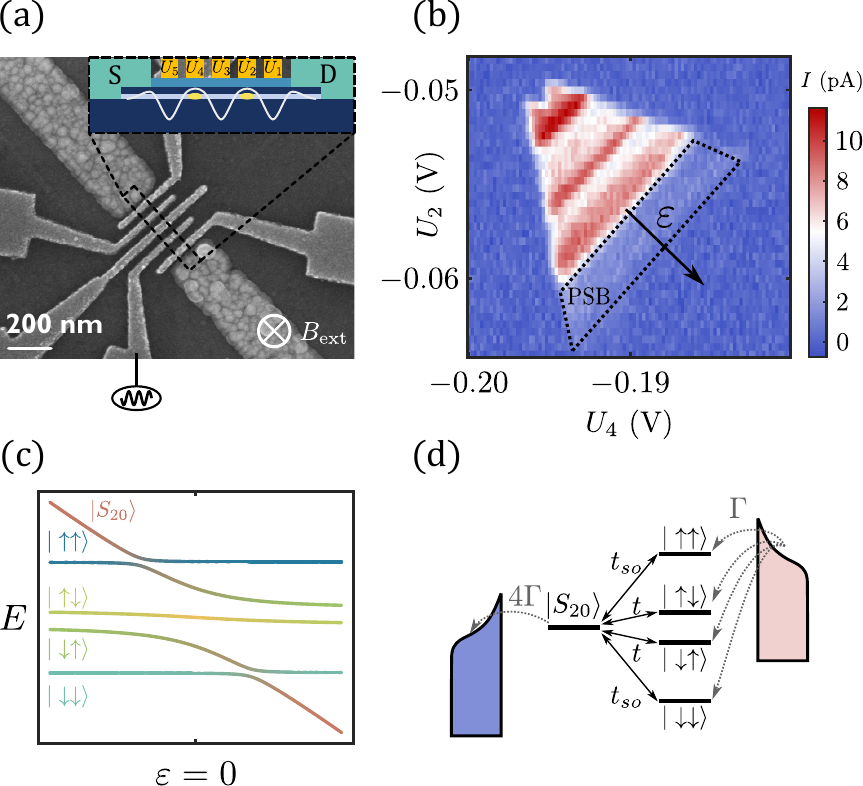}
    \caption{{\bf Scanning electron micrograph of the device and transport current with respect to gate voltages.}
    (a) Scanning electron micrograph of a device of the same structure with the one used in experiment. The nanowire is outlined by the dashed box, of which the cross-section is sketched in the inset. QDs are formed when voltages are applied appropriately.
    The device is exposed to a uniform magnetic field, and microwave drive is applied to gate $U_2$.
    (b) Transport current as a function of gate voltages. The dashed box indicates the current suppression caused by PSB, and detuning $\varepsilon$ is defined accordingly, as shown by the arrow.
    (c) Energy structure of ST system as a function of $\varepsilon$, where the anticrossings suggest the couplings between states.
    (d) Illustration for establishing the transport current. The tunneling events are indicated by the gray dashed arrows, which represent the tunnel of holes in and out of QDs.
    Black solid arrows labeled by $t$ or $t_{so}$ are for the couplings between relative states.
    We assume $\Gamma\gg t,~t_{so}$, and treat the process of jumping in and out by jumping operators $|s_z s_z\rangle\langle S_{20}|$ with $|s_z\rangle = |\uparrow\rangle, |\downarrow\rangle$.
    }
    \label{fig-1}
\end{figure}

Our sample is fabricated in a Ge hut nanowire \cite{Vukusic2017, Li2017, Li2018, Vukusic2018, Gao2020, Xu2020a, Xu2020, Zhang2021, Liu2022}.
The DQD is defined in the nanowire by depositing Aluminum electrodes above, which control the numbers of holes in each dot and the coupling strength in between.  Figure \ref{fig-1} (a) shows a scanning electron micrograph of the device, where the black dashed box indicates the location of the nanowire.
Sketch of the cross-section along the nanowire is presented in the inset, where the DQD potential, from interdot detuning to tunnel barriers, is tuned by gate voltages $U_i$.  In addition, microwave drive is applied to $U_2$ gate when we study driven dynamics.
A uniform magnetic field is applied perpendicular to the nanowire, as shown in Fig.~\ref{fig-1} (a).  The device is cooled in a dilution refrigerator with a base temperature of $200$ mK.

We perform transport measurement through the DQD.  Figure \ref{fig-1} (b) shows the current as a function of gate voltages $U_2$ and $U_4$, which determine the energy levels and thus the occupation of the two dots.  When gate voltages are set in the region outlined by the black dashed trapezoid in Fig.~\ref{fig-1} (b), we observe PSB, when current is blocked due to spin configurations of the holes instead of Coulomb interaction \cite{Kouwenhoven1997, Ono2002, OnoPRL2004, Petta2005}.  In Fig.~\ref{fig-1} (b) we define the interdot detuning $\varepsilon$ along the black solid arrow.  When $\varepsilon$ is tuned deeper into negative beyond the PSB regime, excited orbital states are involved in the transition and lift the PSB, leading to a jump in current.  Based on the width of the PSB region in Fig.~\ref{fig-1} (b), we estimate the magnitude of the excitation energy of the DQD to be $\sim 1 \text{meV}$.

While our DQD is still in the multi-hole regime, in the PSB regime the system is well described by an effective two-hole model near the $(N+2,M+0)$ to $(N+1,M+1)$ charge transition, where $N$ and $M$ refer to the nominal number of core holes in the left and right QDs that do not participate in transport.  More specifically, the low-energy DQD Hamiltonian can be expanded in the basis of five effective two-hole states
$\{|\uparrow\uparrow\rangle,~
|\uparrow\downarrow\rangle,~
|\downarrow\uparrow\rangle,~
|\downarrow\downarrow\rangle,~
|S_{20}\rangle\}$.
Here $|S_{20}\rangle$ denotes a singlet state $|S\rangle = (|\uparrow\downarrow\rangle-|\downarrow\uparrow\rangle)/\sqrt2$ with a charge configuration of $(N+2,M+0)$, while the charge configuration of the other four states has the two valence holes evenly distributed between the dots as $(N+1,M+1)$.
The Hamiltonian in this basis is given by
\begin{equation}
    H =
    \begin{pmatrix}
        \bar E_z &0 &0 &0 & t_{so}\\
        0& \delta E_z & 0& 0& t\\
        0& 0& -\delta E_z & 0& -t\\
        0& 0& 0& -\bar E_z & t_{so}\\
        t_{so} & t & -t & t_{so}& \varepsilon
    \end{pmatrix},
    \label{eq-H}
\end{equation}
where $\varepsilon$ is the detuning defined in Fig.~\ref{fig-1} (b), $\bar E_z=(g_1+g_2)\mu_B B/2$, $\delta E_z=(g_1-g_2)\mu_B B/2$, with $g_{1,2}$ the g-factors of the hole spins in the two QDs, $\mu_B$ the Bohr magneton and $B$ the external magnetic field strength.
Spin-flip tunneling ($t_{so}$) couples $|S_{20}\rangle$ to $|T_\pm\rangle$, while $t$ gives the spin-preserved inter-dot tunnel coupling strength.  
Figure \ref{fig-1} (c) presents the low-energy spectrum as a function of $\varepsilon$ when $B\ne 0$, and Fig.~\ref{fig-1} (d) gives a schematic illustration of possible tunneling events between source/drain and QDs.  Notice that in general $t_{so} \ll t$, even for a hole system like ours, as can be qualitatively seen in the contrast of the current amplitude between the PSB regime and the non-PSB-non-Coulomb-blockade regime shown in Fig.~\ref{fig-1} (b).  Nevertheless, the presence of $t_{so}$ means that in our DQD, PSB is not complete, and a finite leakage current is always present through the DQD \cite{Ono2002, OnoPRL2004}, as shown by the elevated current in the PSB regime compared to the surrounding Coulomb-blocked areas in Fig.~\ref{fig-1} (b).  In the rest of this study, our focus will be on how the leakage current in the PSB regime varies with system parameters and possible driving.

The form of the ST Hamiltonian $H$ is reminiscent of the one for the original observation of CPT \cite{Gray1978}.  In particular, when $B=0$, the system should have three dark states that are orthogonal to $|S_{20}\rangle$, the final state for the transitions here:
$(|\uparrow\uparrow\rangle-|\downarrow\downarrow\rangle)/\sqrt2$,
$(|\uparrow\downarrow\rangle+|\downarrow\uparrow\rangle)/\sqrt2$
and $(\sin\theta|\uparrow\uparrow\rangle-\cos\theta|\uparrow\downarrow\rangle+\cos\theta|\downarrow\uparrow\rangle + \sin\theta|\downarrow\downarrow\rangle)/2$, where $\theta=\arctan(t/t_{so})$.  With these states all in the $(N+1,M+1)$ charge configuration decoupled from the $|S_{20}\rangle$ state in the $(N+2, M)$ configuration, PSB should be strengthened and the leakage current should be suppressed.
In other words, CPT in our DQD should manifest itself as a suppression in the leakage current.

\begin{figure}[tbp]
    \centering
    \includegraphics[width=8.6cm]{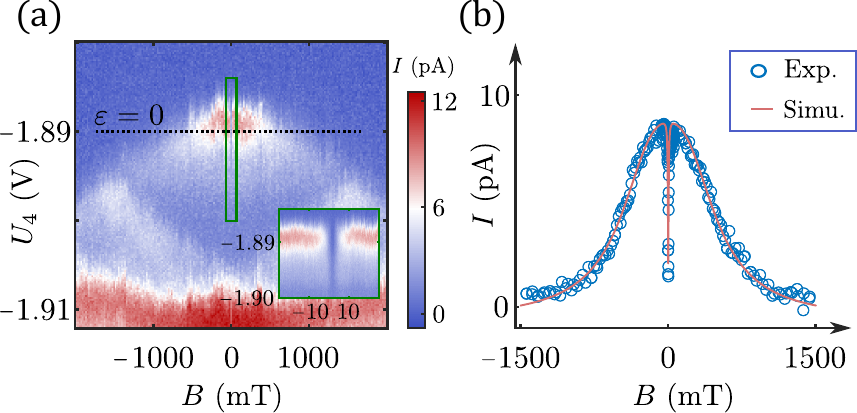}
    \caption{{\bf CPT in ST system without microwave drive.}
    (a) Transport current as a function of strength of magnetic field $B$ and detuning (controlled by $U_4$).
    An abnormal current suppression is observed in the vicinity of $B=0$, which can be attributed to CPT.
    In addition, the zoom-in of the area in the green solid box is shown in the inset, where an apparent dip can be observed.
    (b) Cross-section in the position of $\varepsilon=0$, as indicated by black dashed line in (a).
    Measured current from experiment is represented via scatters, while the red solid curve is for theory fitting.
    The experiment result is well captured by the fitting, from which we estimate the parameters of the system as presented in the main text.
    }
    \label{fig-2}
\end{figure}

To experimentally demonstrate the possible existence of these dark states and therefore CPT, we measured transport current through the DQD versus the applied field $B$ and interdot detuning $\varepsilon$ (controlled by $U_4$), as is shown in Fig.~\ref{fig-2} (a).
This method has been used to characterize mechanisms leading to leakage current through PSB, such as hyperfine interaction, spin-orbit coupling \cite{Nadj-Perge2012, Zhang2021}, \textit{etc.}.
In a significant deviation from previous results, we observe an abrupt and narrow current suppression at zero field, as is shown in Fig.~\ref{fig-2} (a).
The cross-section at $\varepsilon=0$ is measured repeatedly, and the result is shown in Fig.~\ref{fig-2} (b).
The signal is clearly a narrow sharp dip on a broad peak (instead of a superposition of two peaks), which is a distinct feature of CPT.

To understand our observation, we have carried out a detailed theoretical analysis.  Specifically, we model sequential tunneling through the DQD by operator
$T = i\sum_{s_z,s_z'}|s_zs_z'\rangle\langle S_{20}|+h.c.$, where the summation runs over all possible states in the $|(1, 1)\rangle$ configuration, and $h.c.$ stands for Hermitian conjugate \cite{Stoof1996}.
The transport current is then evaluated as $I \propto \text{tr}(T\rho)$, where $\rho$ is the steady-state density matrix of the DQD, obtained from the Lindblad master equation (LME)
\begin{equation}
    \frac{d\rho}{dt} = -\frac{i}{\hbar}[H, \rho] + \sum_k \left(L_k\rho L_k^\dagger - {1\over2}\{L^\dagger_kL_k, \rho\}\right),
    \label{eq-LME}
\end{equation}
with $L_k$ the Lindblad operators representing relaxation and dephasing in the DQD.  In our device, the ST system undergoes a transport cycle of $|S_{20}\rangle\to|(1, 0)\rangle\to|(1, 1)\rangle\to|S_{20}\rangle$ as indicated by the arrows in Fig.~\ref{fig-1} (d).  In practice, the dot-reservoir tunneling rate labeled by $\Gamma$ in Fig.~\ref{fig-1} (d) is in the order of $10$ GHz, which is significantly greater than the interdot tunnel coupling strength $t$ and $t_{so}$.  The steady state population of $|S_{20}\rangle$ is thus very small, since it would rapidly relax to $|(1,0)\rangle$ and the DQD would then quickly reload into one of the $|(1,1)\rangle$ states.  Under this condition, a hole jumping out of the left dot and another jumping into the right dot happen almost simultaneously, with the whole reloading process captured by the Lindblad operator $L_{s_z s_z'}^r=\sqrt\Gamma|s_zs_z'\rangle\langle S_{20}|$, with $\Gamma$ the dot-reservoir tunneling rate and $s_z^{(\prime)} = \uparrow,~\downarrow$.  In addition to this relaxation process, we also consider dephasing of all the energy levels, depicted by the Lindblad operator $L_{i}^d = \sqrt\gamma|i\rangle\langle i|$ with $i$ covering all the basis states of $H$.
The steady-state of the system can be calculated straightforwardly by vectorizing the LME (see Appendix Sec. 1).

To estimate the parameters of our device, we calculate the mean squared error $\text{MSE} = \frac{1}{N}\sum_{i=1}^N|I_i^e-I_i^s|^2$, where the summation runs over all experimental measurements, with $I^e$ and $I^s$ referring to experimental data and numerical simulations, respectively.  Experimental parameters are then estimated by searching for the minimal MSE in a reasonable regime.  Simulation results are shown in Fig.~\ref{fig-2} (b) by the solid line, which agrees very well with the experiment data.  From our simulation results, we estimate the following parameters for our device: $\Gamma\approx 27.5$ GHz, $\gamma \approx 0.655$ GHz, $t_{so}/\hbar\approx 5.32$ GHz and $t/\hbar \approx 2.09$ GHz \cite{Xu2020, Zhang2021, Liu2022}.

\begin{figure}[]
    \centering
    \includegraphics[width=8.4cm]{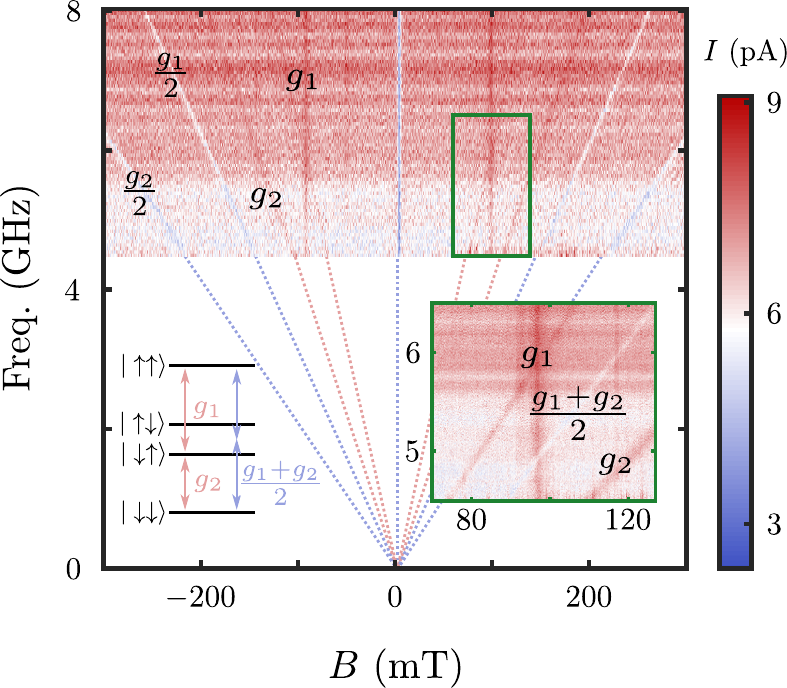}
    \caption{{\bf Odd-even effect and CPT in ST system with longitudinal driving.}
    Transport current as a function of driving frequency and strength of magnetic field is presented, in which enhancement or suppression in transport current can be observed for different harmonics $E_i-E_j=nhf$. Both first- and second-order harmonics are observed in our experiment, the spectrum is relevant to process labeled in the left inset.
    Zoom-in of the area in green solid box is presented in the right inset, in which we see the peaks caused by first-order single-spin rotation and dip caused by second-order two-spin rotation. 
    In addition, the dip in the vicinity of $B\approx 0$ can be attributed to CPT, the width of which is modulated by driving filed.
    }
    \label{fig-3}
\end{figure}

Our experimental measurements, supported by numerical simulations, demonstrate that coherent population trapping can indeed occur in a DQD even without microwave driving.  However, this observed CPT has limited utility due to the degeneracy among the dark states in the five-level system.  
On the other hand, a recent theoretical study \cite{Zhou2022} has shown that a longitudinally driven (at angular frequency $\omega$) ST system can be described by an effective Hamiltonian of a similar structure as conventional CPT systems, near the harmonic resonance condition $E_i-E_j\approx n\hbar\omega$, where $E_{i,j}$ is the energies of the $|(1,1)\rangle$ states.  In such a driven ST system, the resonant driving field allows us to selectively create individual dark states and avoid degeneracy.
This driven CPT could then enable useful applications such as STIRAP.  Moreover, the effective coupling is modulated by driving field, further increasing the tunability of the ST system.  According to our theoretical investigation, such modulation is reflected in two aspects.  Firstly, the modulation cased by a sinusoidal drive leads to an odd-even effect \cite{Stehlik2014, Danon2014, Zhou2022}, i.e., for odd (even) orders of harmonics, the resonance signals are measured as a peak (dip) (see Appendix Sec. 2).  Secondly, the modulation on the effective coupling also inevitably leads to the CPT being modulated at the same time.

We have performed our transport measurement with a microwave applied to the $U_2$ gate, longitudinally driving the interdot detuning of the DQD. In Fig.~\ref{fig-3}, the transport current at $\varepsilon=0$ as a function of driving frequency and external magnetic field strength is presented.  Here we clearly observe current enhancement and suppression caused by the various resonances, indicated by the arrows in the inset of Fig.~\ref{fig-3}.  These resonances include single-spin rotations labeled by the relevant single-dot g-factor $g_{1,2}$, and a two-spin rotation labeled by $(g_1+g_2) / 2$.
From the spectrum, we extract g-factors of the spins in the left and right dot as $g_1 \approx 4.5$ and $g_2 \approx 3.0$.
Our observation here clearly indicates the presence of the odd-even effect for single-spin rotations: for $n = 1$, a resonance yields an enhancement in current, while for $n = 2$, the current is suppressed.

\begin{figure}[]
    \centering
    \includegraphics[width=8.6cm]{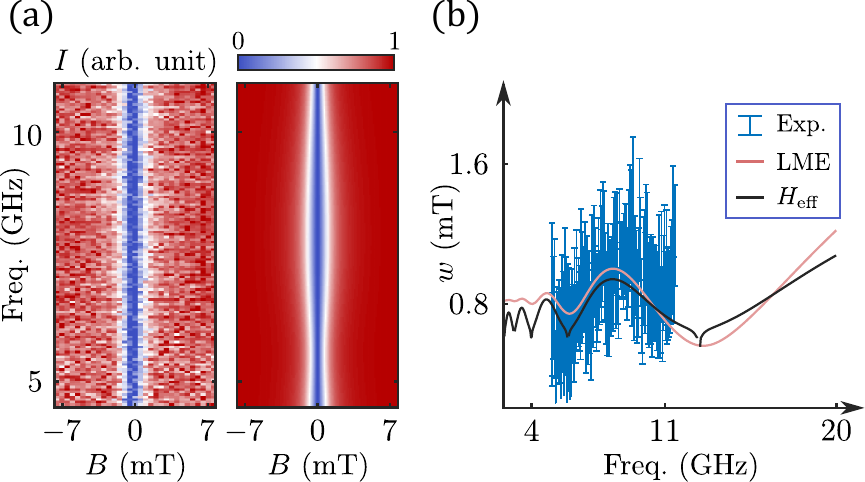}
    \caption{{\bf Modulation of CPT by driving field.}
    (a) Experiment data in the vicinity of $B=0$ mT (left panel) and numerical simulation (right panel). The dip in the middle indicates the modulation on CPT introduced by the longitudinal drive.
    (b) The half width at half maximum $w$ is extracted by fitting with a Lorentzian function.
    $w$ as a function of driving frequency is then presented, with an error bar given by $ 95\%$ confidence interval of the fitting.
    The result fitted from LME ($H_\text{eff}$) is presented by the red (black) solid line, which agrees with the experiment fitting.
    }
    \label{fig-4}
\end{figure}

In addition to the odd-even effect, we also observe a clear dip in current caused by CPT in the vicinity of $B=0$ in Fig.~\ref{fig-3}.
For a driven ST system, CPT can be observed whenever $E_i-E_j = n\hbar\omega$ while $E_i - E_{S} = m\hbar\omega$, where the subscript $i,~j$ represent $(1,1)$ states and $S$ refers to $|S_{20}\rangle$.
This condition can be fulfilled when $\varepsilon =0$ and $B=0$, where a suppression in current is observed as presented in the left panel of Fig.~\ref{fig-4} (a).
We also perform a numerical simulation based on the time-dependent LME with a driving amplitude $A/\hbar = 190$ GHz, and the result is presented in the right panel of Fig.~\ref{fig-4} (a). A consequence of the longitudinal driving we employed is a modulation of the effective couplings when varying the driving frequency.  To reveal this modulation, we extract the half-width-at-half-maximum $w$ as a function of driving frequency $f$ by fitting the current dip at $B = 0$ with a Lorentzian function.
The fitted result is displayed in Fig.~\ref{fig-4} (b) by scatters, where the error bar is given by the $95\%$ confidence interval of the fitting.
Results from the LME and $H_\text{eff}$ are also presented in Fig.~\ref{fig-4} (b) by red and black solid curves.
The $H_\text{eff}$ used here is accurate up to $\mathcal{O}(t_{(so)}J_\nu/\omega)$, in which the coupling strengths are replaced by the modulated ones, i.e., $t_{(so)}\to t_{(so)}J_0(A/\omega)$ (See Appendix Sec. 3).
Figure \ref{fig-4} (b) shows that results from both LME and $H_\text{eff}$ are qualitatively consistent with the experiment data, though $H_\text{eff}$ can be non-convergent when $J_0(A/\omega)=0$ due to the approximations used in its derivation.
The agreement here shows that the clear modulation on CPT originates from the longitudinal driving.  Such a modulation could enable possible applications such as STIRAP in similar systems \cite{Zhou2022}.

It is worth noting that, both odd-even effect and CPT yields a suppression in current, though they are fundamentally different.
Odd-even effect is a direct result of the modulated effective coupling in the form of Bessel function, and is a coherent destruction of tunneling between $|(1,1)\rangle$ and $|S_{20}\rangle$.
In contrast, CPT does not require modulation of the effective coupling.  Instead, it requires that all the relevant energy levels are on resonance, i.e. all diagonal terms in $H_\text{eff}$ are equal, leading to a coherent interference involving all levels.

In summary, we report experimental observation of CPT without driving field based on a five-level ST system for holes in a nanowire device.
The CPT manifests itself as a sharp suppression in the leakage current in the Pauli Spin Blockade regime, and is due to the presence of multiple dark states under certain conditions. 
By applying a longitudinal drive, we have also demonstrated the possibility of selectively creating a dark state and the associated CPT, with the added benefit of an increase in tunability for the effective couplings.  The longitudinal driving also leads to other interesting features to our experimental observations,
from an odd-even effect in the tunnel current to a CPT modulated by the varying driving field frequency.
Our results clearly demonstrate the potential tunability of a longitudinally driven system \cite{Zhou2022}, open up the possibility for STIRAP based quantum gates, and broaden the capacity of the ST system in quantum simulation and quantum computation applications.

\section{acknowledgment}
This work is supported the National Natural Science Foundation of China (Grants Nos. 12074368, 92165207, 12034018 and 61922074), the Anhui Province Natural Science Foundation (Grant No.2108085J03), and the USTC Tang Scholarship. X.H. acknowledges financial support by U.S. ARO through Grant No. W911NF1710257, and this work was partially carried out at the USTC Center for Micro and Nanoscale Research and Fabrication.

\appendix
\section{Appendix}

\subsection{1. Vectorization of Lindblad master equation}
Based on the dissipation model, we calculate the steady-state density matrix by vectorizing the Lindblad master equation given in Eq.~\ref{eq-LME} according to Ref.~\cite{DAriano2000}.
We can think of superoperators as big matrices multiply a big vector $\text{vec}(\rho)$ by mapping $|i\rangle\langle j|\to |i\rangle\otimes|j\rangle$, i.e., the vectorized density matrix as $\text{vec}(\rho) = \sum_{i,j}\rho_{ij}|i\rangle\otimes|j\rangle$.
The vectorized Lindblad master equation can be put into $\text{vec}(\dot\rho) = \hat{\mathcal L}\text{vec}(\rho)$, with $\hat{\mathcal L}$ given by
\begin{equation}
    \begin{aligned}
        &\hat{\mathcal L}=-{i\over\hbar}(I\otimes H-H^T\otimes I)+\\
        &\sum_k\left[L_k^*\otimes L_k-{1\over2}I\otimes L_k^\dagger L_k- {1\over2}(L_k^\dagger L_k)^T\otimes I   \right].
    \end{aligned}
    \label{eq-vLME}
\end{equation}
Proceeding along these lines, the master equation is transformed into a matrix-vector equation.
Notice that $\text{tr}(I^\dagger \rho)=\text{vec}(I)^\dagger\text{vec}(\rho)\equiv 1$, we have $\text{vec}(\dot\rho)=\text{vec}(I)^\dagger\hat{\mathcal L}\text{vec}(\rho)=0$.
This powerful result shows that any trace-preserving $\hat{\mathcal L}$ must have a zero eigenvalue, of which the left eigenvector will be identity whereas the right eigenvector be the steady-state density matrix.
This result provides us a convenient approach to the steady state.

\subsection{2. Odd-even effect induced by strong driving}

The odd-even effect results from the modulation on effective couplings, which is caused by longitudinal sinusoidal drive.
Here, we concisely show that the odd-even effect can be illustrated based on the effective Hamiltonian $H_\text{eff}$ according to our previous work \cite{Zhou2022}.
Take the resonance $\bar E_z-\delta E_z=n\omega$ as an example.
Transition from $|\uparrow\uparrow\rangle$ to $|S_{20}\rangle$, which causes transport current, can be determined by coupling strength.
However, in the driven system, the energy levels are dressed by the driving field and are coupled to each other, the transition is determined by a perturbated coupling $r$ according to our previous work.
For $\bar E_z-\delta E_z=n\omega$, the relevant effective coupling strength $r$ is given by
\begin{equation}
    r\propto\sum_{m\ne N}\frac{t_{so}t^2}{m\omega+\delta E_z-\varepsilon_0}(J_{m-n}J_mJ_N-J_m^2J_{N-n}),
\end{equation}
where $n$ denotes the different harmonics while $N$ is the integer closest to $\varepsilon/\omega$.
According to the asymptotic behavior of Bessel function
\begin{equation}
    J_n(x)\approx \sqrt{\frac{2}{\pi x}}\cos[x-(2n+1)\pi/4],\quad x\gg n,
\end{equation}
when $A/\omega$ is sufficiently great compared to $\varepsilon/\omega$, we obtain $r\approx 0$ for any even $n$.
In other words, the state transition is coherently destructed for an even $n$, while that for an odd $n$ is not, which manifests as an odd-even effect.

\subsection{3. Effective Hamiltonian}

The external driving can be described by $\varepsilon = \varepsilon_0 + A\cos\omega t$.
In the $H_\text{eff}$, $t$ and $t_{so}$ are modulated in the form of $t_{(so)}J_\nu(A/\omega)$, and there are synthetic couplings between any two $(1,1)$ states, which is in the magnitude of $\mathcal O(t_{(so)}^2J_\nu^2/\omega)$.
Take the resonance $\bar E_z-\delta E_z=n\omega$ as an example.
When the system is near resonance, the $H_\text{eff}$ is approximately a three-level model in the basis of $\{|\uparrow\uparrow\rangle,~|\uparrow\downarrow\rangle,~|S_{20}\rangle\}$.
The effective Hamiltonian is expressed as
\begin{equation}
    H_\text{eff}=
    \begin{pmatrix}
        g_2\mu_BB-n\omega & \xi & t_{so}J_{N-n}(\frac{A}{\omega})\\
        \xi & \Delta & tJ_N(\frac{A}{\omega})\\
        t_{so}J_{N-n}(\frac{A}{\omega}) & tJ_N(\frac{A}{\omega}) & \varepsilon_0-\delta E_z-N\omega-\Delta
    \end{pmatrix},
\end{equation}
where $J_\nu(x)$ is the Bessel function of the first kind, $\xi$ and $\Delta$ are induced by driving field, given by
\begin{equation}
    \xi = \sum_{m\ne N}\frac{t_{so}tJ_mJ_{m-n}}{2(\delta E_z-\varepsilon_0+m\omega)} + \sum_{m\ne N-n}\frac{t_{so}tJ_mJ_{m+n}}{2(\bar E_z-\varepsilon_0+m\omega)}
\end{equation}
\begin{equation}
    \Delta =\sum_{m \ne N}\frac{(t J_m)^2}{\delta E_z-\varepsilon_0+m\omega}.
\end{equation}

Moreover, we study the CPT near $B=0$ with the effective model as exhibited in Fig~\ref{fig-4} (b). The effective model $H_\text{eff}$ used here is a five-level model expressed as
\begin{equation}
    H_\text{eff} =
    \begin{pmatrix}
        \bar E_z &0 &0 &0 & \tilde t_{so}\\
        0& \delta E_z & 0& 0& \tilde t\\
        0& 0& -\delta E_z & 0& -\tilde t\\
        0& 0& 0& -\bar E_z & \tilde t_{so}\\
        \tilde t_{so} & \tilde t & -\tilde t & \tilde t_{so}& \varepsilon
    \end{pmatrix},
\end{equation}
with $\tilde t_{(so)} = J_0(A/\omega)t_{(so)}$.
The $H_\text{eff}$ is accurate up to $\mathcal{O}(t_{(so)}J_\nu/\omega)$ and is qualitatively consistent with experiment data, though it fails to capture the CPT when $J_0(A/\omega)\approx0$.
In addition, the original Hamiltonian $H$ and effective Hamiltonian $H_\text{eff}$ are expressed in different frames, thus the dissipation rates are different, which is another reason for the difference between black and red curves in Fig~\ref{fig-4} (b).

\bibliography{reference.bib}

\end{document}